\documentclass[fleqn,twoside]{article}
\usepackage{espcrc2}
\usepackage{graphicx,epsfig,psfig}
\usepackage[figuresright]{rotating}

\newcommand{\AmS}{{\protect\the\textfont2
  A\kern-.1667em\lower.5ex\hbox{M}\kern-.125emS}}
\newcommand{\cd}{\makebox[0.08cm]{$\cdot$}}

\newcommand{\sla}{\not\!}

\title{Non-perturbative renormalization in Light Front Dynamics with Fock space truncation}

\author{J.-F. Mathiot\address{Laboratoire de Physique Corpusculaire, Blaise Pascal University\\24 avenue des Landais, F-63177 Aubi\`ere Cedex, France},
        V.A. Karmanov\address[lebedev]{Lebedev Physical Institute, Leninsky Prospekt 53, 119991 Moscow, Russia} \thanks{Supported in part by the Russian Foundation for Basic Research grant 05-02-17482-a},
        and
    A.V. Smirnov\addressmark[lebedev] $^*$ }

\begin{document}
\begin{abstract}
Within the framework of the Covariant formulation of  Light-Front
Dynamics, we develop a general non-perturbative renormalization
scheme based on the Fock decomposition of the state vector and its
truncation. The explicit dependence of our formalism on the
orientation of the light front is essential in order to analyze
the structure of the counterterms and bare parameters needed to
renormalize the theory. We present here a general strategy to
determine the dependence of these quantities on the Fock sectors.
We apply our formalism to QED for the two-body (one fermion and
one boson) truncation and recover analytically, without any
perturbative expansion,  the renormalization of the electric
charge according to the requirements of the Ward Identity.
\vspace{1pc}
\end{abstract}
\maketitle
\section{INTRODUCTION}
The knowledge of the hadron properties within the framework of QCD
is one of the main issue in strong interaction physics.  Several
approaches have been pursued in the last twenty years, in
particular lattice gauge calculations.  Among the alternatives to
these calculations, Light-Front Dynamics (LFD) is of particular
interest \cite{bpp}.  It has proven successful in many
phenomenological applications involving few-body systems in
particle and nuclear physics. However, the application of LFD to
field theoretical calculations is still in its infancy. The main
issue to be solved is the renormalization procedure. In
perturbative calculations, the renormalization of the electron
self-energy in QED, in standard LFD, is already non-trivial in the
sense that it involves non-local counterterms.  This unpleasant
feature is however a direct consequence of the choice of a
preferential direction, the $z$ axis, in the determination of the
quantization plane.  This can be well understood in the Covariant
formulation of Light-Front Dynamics (CLFD) \cite{cdkm,kms}. In
this formulation, the state vector is defined on the light-front
plane given by the equation $\omega \cd x=0$, where $\omega$ is
the four-vector with $\omega^2=0$. The particular case where
$\omega=(1,0,0,-1)$ corresponds to standard LFD. We shall show in
this study how one should determine, in a true non-perturbative
way, the renormalization condition for systems composed of one
fermion and at most $N-1$ bosons. The first calculation of the
purely scalar system in CLFD, for $N=3$, has been done in
\cite{bckm}. Non-perturbative calculation in a gauge theory (for
$N=2$) is given in \cite{kms,Br_etal} and in Yukawa theory (for
$N=3$) -- in \cite{BHMc}.

\section{EQUATION OF MOTION IN TRUNCATED FOCK SPACE}
Our starting point is the general eigenstate equation for the state vector $\phi(p)$:
\begin{equation}\label{p2}
\hat{P}^2\ \phi(p)=M^2\ \phi(p) \ ,
\end{equation}
where $M$ is the mass of the physical state. The momentum operator $\hat{P}_{\mu}$ is decomposed into two
parts: the usual free one  and the interacting one given by
\begin{eqnarray}\label{pint}
\hat{P}^{int}_{\mu}&=&\omega_{\mu}\int H^{int}(x)\delta(\omega\cd x)
\ d^4x\\
&=&\omega_{\mu}\int_{-\infty}^{+\infty}
\tilde{H}^{int}(\omega\tau)\frac{d\tau}{2\pi} \ ,
\end{eqnarray}
where $\tilde{H}^{int}$ is the Fourier transform of
the interaction Hamiltonian $H^{int}(x)$ .
The state vector  $\phi(p)$ is then decomposed in Fock components $\phi_i$.
For convenience, it will be more appropriate to work with the vertex function defined by
\begin{equation}
\Gamma_i = 2(\omega \cd p) \tau \phi_i=(s-M^2) \phi_i\ ,
\end{equation}
where $s$ is the square of the invariant mass of the $i$-particle system.
The general equation of motion can thus be written, after some algebraic manipulation, as
\begin{equation}\label{calG}
{\cal G}(p)=-\int\tilde{H}^{int}(\omega\tau)\ \frac{1}{\hat{\tau}}\ {\cal G}(p)\
\frac{d\tau}{2\pi}\ ,
\end{equation}
where ${\cal G}(p)$ is defined by:
\begin{equation}
{\cal G}(p)=2(\omega \cd p)\  \hat{\tau}\  \phi (p)\ .
\end{equation}
The operator $\hat{\tau}$ acts on a given Fock component $\phi_i$ to give $\tau \phi_i$. ${\cal G}(p)$ is thus
decomposed in an infinite sum of the vertex functions $\Gamma_i$.

The interacting Hamiltonian $\tilde{H}^{int}(\omega \tau)$ relates the component $n$ with components $n,n\pm 1, n\pm 2,...$ depending on the system under consideration. It is expressed in terms of bare quantities like the bare coupling constant $g_0$, and/or physical quantities like the mass $m$ of the constituent fermion with a mass counterterm $\delta m$. In order to calculate physical observables, these bare quantities and counterterms should be fixed from physical conditions, in the limit where $M \to m$. If we restrict ourselves to QED in order to compare with well established results in perturbation theory, the mass counterterm is fixed to get $P^2 = M^2$ for the physical state, and $g_0\equiv e_0$ is fixed to get the physical charge of the electron, with $\alpha = \frac{1}{137}=\frac{e^2}{4\pi}$.\\

What happens if we truncate the Fock expansion? We shall call $N$ the total number of Fock components considered, and $n$ a particular Fock component with  $n$ particles in the intermediate state (one fermion and $n-1$ bosons). \\

{\it i)} First Eq.(\ref{calG}) should be modified for $n=N$ or less since $\tilde {H}^{int}\  {\cal G}$ should not connect states with $n > N$. This is done "by hand" in the coupled eigenvalue equations for the vertex parts $\Gamma_n^{(N)}$. We have explicitly mentioned the dependence  of $\Gamma$ on $N$ by a superscript.\\

{\it ii)}
When the bare coupling constant $g_0$ and mass counterterm $\delta m$ are fixed from physical observables, this is done within the approximation of Fock truncation. In each intermediate state, for the calculation of a given Fock component, the total number, $l$, of particles in which a given state can fluctuate should be less or equal to $N$.

\begin{figure}[btph]
\begin{center}
\includegraphics[width=18pc]{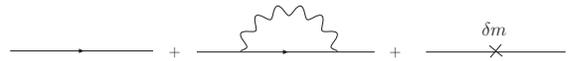}
\caption{Renormalization of the bare fermion propagator \label{self}}
\end{center}
\end{figure}

Let us look for example at the renormalization of the fermion propagator in first order perturbation theory. The three contributions to the physical fermion propagator are indicated in Fig.\ref{self}. These are from left to right the free propagator, the self energy contribution, and  the contribution from the mass counterterm $\delta m$. The sum of these three contributions should be equal, at $p^2=m^2$, to the free propagator. This fixes $\delta m = -g_0^2\Sigma(p^2=m^2)$. This equation links two diagrams with two different Fock components. To keep track of the number of Fock component the counterterm is associated to (here $2$), we use the superscript $(l)$, and note $\delta m^{(l)},g_0^{(l)}$. We shall see in the following section how this dependence should be calculated in practice. \\

{\it iii)} The equation of motion for the vertex parts involves a bare coupling constant. This bare coupling constant is determined to get a physical condition, in a given approximation, i.e. within the Fock truncation. However, it cannot be identified with the usual bare coupling constant which is determined from external meson coupling like the electromagnetic form factor in the case of QED. In the determination of the state vector, the exchanged meson is indeed not an external particle and thus should participate to the counting rule of the total number of particles either in the initial or final state.\\

\begin{figure}[btph]
\begin{center}
\includegraphics[width=18pc]{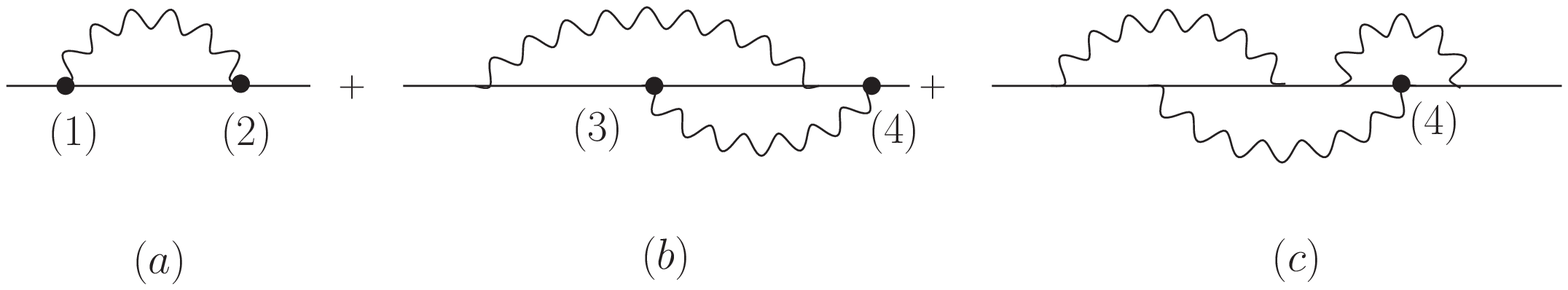}
\caption{Second order radiative corrections to the self-energy of a fermion \label{radc}}
\end{center}
\end{figure}
\begin{figure}[btph]
\begin{center}
\includegraphics[width=10pc]{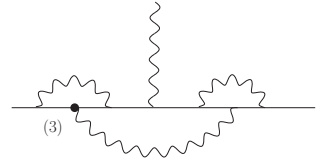}
\caption{Second order correction to the electromagnetic coupling.\label{elm2}}
\end{center}
\end{figure}

To illustrate this fact, let us write down for instance the first contributions to the self energy involving at most two mesons in flight. These are indicated in Fig.\ref{radc}. The vertices $(1)$ and $(2)$ in Fig.\ref{radc}.(a) involve bare coupling constants in order to cancel divergences arising in Fig.\ref{radc}.(b) and (c). However, the vertex $(3)$ in Fig.\ref{radc}.(b) should not be corrected in the same way as the vertex $(4)$ since the vertex $(3)$ can not be modified anymore by radiative corrections, while the vertex $(4)$ can, as indicated in Fig.\ref{radc}.(c). The vertex $(3)$ would however be corrected if one takes as bare coupling constant the one fixed from the calculation of the electromagnetic  form factor, as indicated in Fig.\ref{elm2}. We shall denote by $\bar g_0^{(l)}$ the "Amputated" Bare Coupling Constant (ABCC) to be used in the equation of motion.

The strategy to calculate $\delta m^{(l)}$ and $\bar{g_0}^{(l)}$ is thus the following. Any calculation of the state vector to order $N$ involves $\delta m^{(l)}$ and $\bar{g_0}^{(l)}$ for $l=1,2,...N$. These quantities are however not arbitrary. They are just the realization of the only two quantities $\delta m$ and $\bar{g_0}$ for successive approximate calculations. One should fix them by calculating the state vector for $N=1,2,...N$ successively. The case $N=1$ is trivial since $\delta m^{(1)}=\bar{g_0}^{(1)}=0$ because the state vector corresponds just to a free fermion. The case $N=2$ is calculated in section $4$ for QED.
\section{SPECIFIC LIGHT FRONT COUNTERTERMS}
We have seen in \cite{kms,mks} that the 2-Point Green Function (2PGF) is in general dependent on $\omega$. This comes from the general $\omega$-dependence of the self-energy according to:
$$
\nonumber {\cal M}^{(2)}(p) = g^2\left[{\cal A}(p^2)+{\cal B}(p^2)\frac{\sla{p}}{m}+
{\cal C}(p^2)\frac{m\sla{\omega}}{\omega\cd p}\right]\ .
$$
We have thus exhibited a new counterterm $Z_\omega$ so that the physical 2PGF does not depend on $\omega$. We found that it corresponds also to the counterterm needed to get a two-body component independent of $\omega$ at $s=M^2$. According to the previous discussion, we shall denote this counterterm by $Z_\omega^{(l)}$. Note that for appropriatly chosen regularization schemes, one may find ${\cal C}=0$. Such regularization schemes (Pauli-Villars substraction for instance) are thus clearly favoured.

For the same reason, one should also consider in principle new $\omega$-dependent vertex counterterms, denoted by
\begin{equation} \label{gw}
\delta g_\omega = \sum_i\delta g_{\omega,i}\  O_i(\omega)\ .
\end{equation}
with appropriate structures $O_i(\omega)$ depending on the type of vertices (scalar or vector). These counterterms are needed to make sure that the physical 3-point Green function, like the electromagnetic form factor for instance, is independent of $\omega$. These counterterms should also depend on $(l)$.

For the calculation we consider in this work, we do not need to discuss higher n-point Green function.
\section{THE CASE OF QED FOR N=2} \label{QED}
We follow here very closely the definitions already presented in \cite{kms}.

\subsection{Decomposition of the state vector}
In the two-body Fock truncation we consider in this section, the state vector is written as
\begin{equation}
\vert p \sigma \rangle \equiv \vert 1 \rangle + \vert 2 \rangle
\end{equation}
with
\begin{eqnarray*}
\vert 1 \rangle&=&(2\pi)^{3/2}\sum_{\sigma'}\int
\phi_{1,\sigma\sigma'}(p_1,p,\omega\tau_1) a^{\dag}_{\sigma'}({\bf
p}_1)|0\rangle
\nonumber\\
&\times&\delta^{(4)}(p_1-p-\omega\tau_1)2(\omega\cd p)
\frac{d\tau_1 d^3p_1}{(2\pi)^{3/2}\sqrt{2\varepsilon_{p_1}}}
\nonumber\\
\vert 2 \rangle&=&(2\pi)^{3/2}\sum_{\sigma'}\int \phi_{2,\sigma\sigma'}
(k_1,k_2,p,\omega\tau_2) \nonumber\\
&\times& a^{\dag}_{\sigma'}({\bf k}_1)c^{\dag}({\bf
k}_2)|0\rangle
\delta^{(4)}(k_1+k_2-p-\omega\tau_2)\nonumber\\
&\times& 2(\omega\cd p)
\frac{d\tau_2
d^3k_1d^3k_2}{(2\pi)^{3/2}\sqrt{2\varepsilon_{k_1}}(2\pi)^{3/2}
\sqrt{2\varepsilon_{k_2}}}\ , \nonumber
\end{eqnarray*}
and the normalization condition is given by
\begin{equation} \label{norma}
\langle p' \sigma ' \vert p \sigma \rangle = 2 \varepsilon_p\  \delta_{\sigma, \sigma '} \delta(\bf{p} - \bf{p'}) \ .
\end{equation}
The one-body  $\phi_{1,\sigma\sigma'}(p_1,p,\omega\tau_1)$ and two-body $\phi_{2,\sigma\sigma'}
(k_1,k_2,p,\omega\tau_2)$ wave functions are thus decomposed as follows:
\begin{eqnarray}
\phi_{1,\sigma\sigma'}&\equiv& \frac{\bar{u}_{\sigma'}(p_1)\ \Gamma_1\ u_{\sigma}(p)}{s-M^2} = \phi_1\  \bar{u}_{\sigma'}(p_1)u_{\sigma}(p)\nonumber \\
\phi_{2,\sigma\sigma'}&\equiv& \frac{\bar{u}_{\sigma'}(k_1)\ \Gamma_2\ u_{\sigma}(p)}{s-M^2} \nonumber \\
&=&\frac{1}{s-M^2}\ \bar{u}_{\sigma'}(k_1) \left[ a_2  + b_2\frac{m \sla \omega }{\omega \cd p}\right] u_{\sigma}(p)  \nonumber \ .
\end{eqnarray}
Note that due to our explicit covariant formulation we can easily
decompose the two-body component into its two spin components
$a_2$ and $b_2$ \cite{cdkm}. Since the state vector is defined on
the light front plane $\omega \cd x=0$, it should therefore depend
explicitly on $\omega$. The components $\phi_1, a_2, b_2$ are then
determined from the equation of motion. In the approximation where
one considers only one fermion and one boson state, these
components are constants. This is however not the case in general.

In order to make a clear connection with the usual normalization factor $Z_2^{(2)}$ we shall define
\begin{equation}
\sqrt{Z_2^{(2)}} \equiv 2m\ \phi_1 \ .
\end{equation}
The normalization factor  $Z_2^{(2)}$ is fixed from the condition (\ref{norma})
%
\subsection{Solution of the equation of motion}
We shall only recall here the results obtained in \cite{kms}, with
the identification $\delta m \to \delta m^{(2)},\ g \to \bar
g_0^{(2)},\ Z_\omega \to Z_\omega^{(2)}$. The counterterm $\delta
m^{(2)}$ is fixed from the equation for $\Gamma_2$ in order to get
an $\omega$-independent two-body wave function at $s=M^2$, i.e.
$b_2 = 0$. This gives
\begin{equation}
\delta m^{(2)} = \left[ \bar g_0^{(2)} \right]^2 (A + B + C)\ ,
\end{equation}
while $Z_\omega^{(2)}$ is determined by solving the equation for $\Gamma_1$, so that
\begin{equation}\label{Z2}
Z_\omega^{(2)} = \left[ \bar g_0^{(2)} \right]^2 C \ .
\end{equation}
The two body component $a_2$ is thus given by
\begin{equation}
a_2 = 2m\  \phi_1\ \bar g_0^{(2)} = \bar g_0^{(2)}\  \sqrt{Z_2^{(2)}}\ .
\end{equation}
The quantities $A,B,$ and $C$ are given by ${\cal A}, {\cal B,}$
and ${\cal C}$ at $p^2=M^2$. Note that to simplify notations, we
include in $C$ the contribution from the boson loop with the
contact fermionic interaction \cite{kms}. With this, one can
completely determine the normalized wave function and find from
the normalization condition (\ref{norma}):
\begin{equation} \label{Z2g0}
Z_2^{(2)}=\frac{1}{1+\left[\bar g_0^{(2)}\right]^2 I(L)}\ ,
\end{equation}
with $I(L)$  given by
$$
\nonumber I(L)=\frac{1}{2(2\pi)^3} \int \frac{Tr[(\sla k + m)(\sla p +m)]}{(s-M^2)^2} \frac{d^3k}{2\varepsilon_k (1-x)}
$$
where $k$ is the photon momentum, and $L$ is a cut-off needed to give a meaning to the integral. For our purpose, we do not need to precise the regularization scheme we use. $I(L)$ is a logarithmically divergent quantity as a function of $L$.
\subsection{Determination of the amputated bare coupling constant}
In the spirit of our previous choice of $b_2=0$ at $s=M^2$, we can define the ABCC $\bar g_0^{(2)}$, so that the component $a_2$ at $s=M^2$ is identified with the physical coupling constant:
\begin{equation} \label{a2}
a_2(s=M^2) = \bar g_0^{(2)} \sqrt{Z_2^{(2)}} \equiv g\ .
\end{equation}
It is a well defined, non-perturbative, condition which is very convenient to impose in any numerical calculation (apart from the fact that we should extrapolate the component to the non-physical point $s=M^2$; this is here trivial since $a_2$ is a constant).

\begin{figure}[btph]
\begin{center}
\includegraphics[width=14pc]{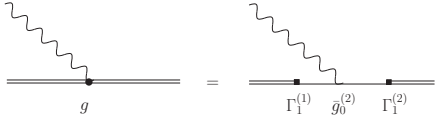}
\caption{Calculation of the "amputated" bare coupling constant for $N=2$. To this order, there is only one contribution to the electromagnetic amplitude {\it photon + fermion} $\to$ {\it fermion} \label{abcc2}}
\end{center}
\end{figure}

We can note that according to our discussion in section 1, $\bar g_0^{(2)}$ can also be extracted from the calculation of the meson-nucleon vertex. In the $N=2$ approximation, it is given by the contribution of Fig.\ref{abcc2}, with the result:
\begin{equation} \label{g0b}
g=\sqrt{Z_2^{(2)}} \bar g_0^{(2)} \sqrt{Z_2^{(1)}}=\sqrt{Z_2^{(2)}} \bar g_0^{(2)}\ .
\end{equation}
We recover exactly Eq.(\ref{a2}). In the r.h.s. of
Fig.\ref{abcc2}, the final state can fluctuate into a fermion plus
a boson. It is thus given by $\Gamma_1^{(2)}$. This is however not
the case for the initial state since there is already a boson in
flight, so that it is given by $\Gamma_1^{(1)}$, $i.e.$ a free
fermion state with $Z_2^{(1)}=1$. Since the contribution from the
r.h.s. of Fig.\ref{abcc2} is already $\omega$-independent, we
should not consider any $\omega$-dependent counterterms, so that
$\delta \bar g_{\omega,i}^{(2)}=0$. This validates the use of an
$\omega$-independent coupling constant in the $N=2$ calculation of
\cite{kms}.

With this definition of $\bar g_0^{(2)}$, and with Eq.(\ref{Z2g0}), we have
\begin{equation}
g=\frac{\bar g_0^{(2)}}{\sqrt{1+\left[ \bar g_0^{(2)}\right]^2  I(L) }}\ ,
\end{equation}
which gives
\begin{equation} \label{g0b22}
\left[ \bar g_0^{(2)} \right]^2=\frac{g^2}{1-g^2I(L)}\ ,
\end{equation}
and thus
\begin{equation} \label{Z2g}
Z_2^{(2)} = 1-g^2I(L)\ .
\end{equation}
We would like to emphasize here that we did not do any perturbative expansion in order to express $Z_2^{(2)}$ in Eq.(\ref{Z2g}) starting from Eq.(\ref{Z2g0}). The first one is expressed in terms of the physical coupling constant $g$ while the second one is a function of the ABCC $\bar{g_0}$.

\subsection{Determination of the bare coupling constant}
As already explained in \cite{km}, there are three
$\omega$-dependent structures that we can think of in order to
expand the electromagnetic form factors in QED, on top of the two
physical ones $F_1$ and $F_2$. We recall here these three
structures:
\begin{eqnarray}
O_1^\rho&=&\left(\frac{m\sla \omega}{\omega \cd p}-\frac{1}{1+\eta}\right) \frac{(p+p')^\rho}{m} \ ,\\
O_2^\rho&=&\frac{m}{\omega \cd p}\omega^\rho \ ,\\
O_3^\rho&=&\frac{m^2}{(\omega \cd p)^2}\sla \omega \omega^\rho
\end{eqnarray}
with $\eta=Q^2/4m^2$, where $Q^2=-q^2$. The most general decomposition of the total amplitude thus writes
\begin{eqnarray} \label{gamr}
\bar u\   \Gamma^\rho\  u&=&g\ \bar u \left[F_1(Q^2) \gamma^\rho + F_2(Q^2) \frac{i}{2m}\sigma^{\rho \nu} q_\nu \right. \nonumber \\
&+&\left. \sum_{i=1}^{3} \ B_i(Q^2) \  O_i^\rho  \right] u \ .
\end{eqnarray}
Each  form factor can be easily extracted from $ \Gamma^\rho$
according to the procedure detailed in \cite{km}.

The Bare Coupling Constant (BCC) $g_0^{(l)}$ should thus be fixed to get the physical coupling constant defined by $g= g\ F_1(0)$, i.e. $F_1(0)=1$, while the $\omega$-dependent counterterms $\delta g_{\omega,i}^{(2)}$ defined in Eq.(\ref{gw}) should be fixed to get an $\omega$-independent amplitude, if needed.
\begin{figure}[tbph]
\begin{center}
\includegraphics[width=14pc]{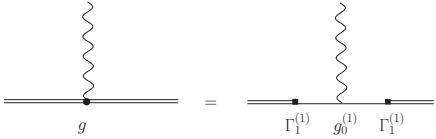}
\caption{Calculation of the bare coupling constant for $N=1$ \label{bcc1}}
\end{center}
\end{figure}

For completness, we shall first examine the trivial calculation for $N=1$. In this case, the form factor is given by Fig.\ref{bcc1} so that
\begin{eqnarray}
g_0^{(1)}&=&g \label{g0g}\ ,\\
\delta g_\omega^{(1)}&=&0\ ,
\end{eqnarray}
since the contribution from the r.h.s. of Fig.\ref{bcc1} is independent of $\omega$. \\
\begin{figure}[btph]
\begin{center}
\includegraphics[width=18pc]{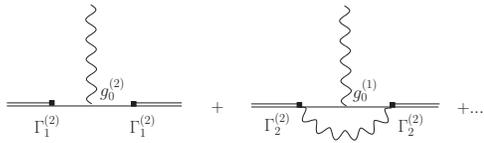}
\caption{Contributions to the form factor in the two-body approximation. The dots indicate the contribution from fermionic contact interactions\label{ff2}}
\end{center}
\end{figure}
For the case $N=2$, the contributions to the form factor are indicated on Fig.\ref{ff2}.  Direct calculation of the contributions from fermionic contact interactions  gives no contribution at $Q^2=0$ to the $\omega$-independent contribution $F_1$.

By definition, the vertex counterterms $\delta g_\omega^{(2)}$ are chosen  to cancel the $\omega$ dependent contributions arising from the diagrams in Fig.\ref{ff2}, so that $B_{1,2,3}$ in Eq.(\ref{gamr}) are equal to zero.  We do not detail here their calculation. They depend directly on the regularization scheme used in the evaluation of the loop diagrams.

One is thus left with one equation for the $\omega$-independent contribution which writes:
\begin{eqnarray} \label{g0}
g&=&\sqrt{Z_2^{(2)}} g_0^{(2)}\sqrt{Z_2^{(2)}} \nonumber \\
&+& \sqrt{Z_2^{(2)}} F^{(2)}(Q^2=0,L)g_0^{(1)}\sqrt{Z_2^{(2)}} \ ,
\end{eqnarray}
where the factor $F^{(2)}$ is the contribution coming from the $\omega$-independent part of the loop diagram  in Fig.\ref{ff2}.(b) for $N=2$.

The difference between $g_0^{(2)}$ in Eq.(\ref{g0}) and $\bar
g_0^{(2)}$ in Eq.(\ref{g0b}) lies in the fact that the initial
one-body state cannot fluctuate into a two-body part in the
calculation of  $\bar g_0^{(2)}$, so that the one body ($n=1$)
component should correspond to $N=1$ and not to $N=2$ like in the
form factor calculation. One can easily check that for QED
\cite{dkm}, we have
\begin{equation}
F^{(2)}(Q^2=0,L)=\left[ \bar g_0^{(2)} \right]^2 I(L)\ ,
\end{equation}
so that Eq.(\ref{g0}) writes with (\ref{g0b}) and (\ref{g0g}):
\begin{equation}
g=\left[1-g^2I(L)\right]  g_0^{(2)} + g^3 I(L)\ ,
\end{equation}
which gives finally
\begin{equation} \label{WI}
g_0^{(2)}=g\ .
\end{equation}
This is in complete agreement with the Ward Identity. This originates from two particular features of our strategy: the ABCC from (\ref{g0b}) which implies (\ref{Z2g}) instead of (\ref{Z2g0}), and the definition of the Fock-dependent BCC in (\ref{g0}) and (\ref{g0g}).

\subsection{Calculation of the form factor} \label{calff}
Let us now investigate the general form of the electromagnetic form factor for any $Q^2$. It is given by the contributions shown in  Fig.\ref{ff2}. The physical form factors are thus extracted from our general decomposition (\ref{gamr}).

The $\omega$-independent contributions from Fig.\ref{ff2} have been already calculated in \cite{dkm}. If we restrict ourselves for instance to the $F_1$ form factor, we get
\begin{equation}\label{ge}
gF_1(Q^2)=g_0^{(2)}Z_2^{(2)} + g_0^{(1)} Z_2^{(2)}F^{(2)}(Q^2,L)
\end{equation}
with
\begin{equation}
F^{(2)}(Q^2,L)=\left[ \bar g_0^{(2)} \right]^2 {\bar F}^{(2)}(Q^2,L)
\end{equation}
so that we have immediately with (\ref{g0b22}), (\ref{Z2g}), (\ref{g0g}), and (\ref{WI}):
$$
F_1(Q^2)=1+g^2\left[{\bar F}^{(2)}(Q^2,L)- {\bar F}^{(2)}(Q^2=0,L)\right]
$$
We recover here exactly the perturbative calculation, as it is natural in the two-body approximation. Note that this result is obtained without any perturbative expansion.

\section{CONCLUDING REMARKS AND PERSPECTIVES}
We have presented in this contribution a systematic strategy to calculate physical observables for fermionic systems composed of a fermion and $N-1$ bosons. This implies to implement a renormalization scheme in a non-perturbative framework. Within the covariant formulation of Light Front Dynamics, we have shown how to fix the mass counterterm and bare coupling constant of the elementary Hamiltonian in a consistent way. As a check of our formalism, we treat the case of QED in the two body approximation. We have been able to recover, for the first time, the standard renormalization of the electromagnetic charge according to the Ward Identity (i.e. the bare charge should be equal to the physical one), without any perturbative expansion. This shows that no divergences are left uncancelled.

Our results have been made possible because of two important features of our formalism:\\

{\it i)} First we should extract the physical part of the two-body wave function at $s=M^2$ in order to define the physical coupling constant to be used in the calculation of the state vector. This part is explicit in our formalism since it should be independent of $\omega$.\\

{\it ii)} Since we truncate the Fock expansion, the physical content of the bare coupling constant to be used in the state vector should be different from the bare coupling constant calculated from external meson coupling (by looking at the electromagnetic form factor for instance if we deal with QED). These two couplings are of course the same in the limit of infinite Fock decomposition.\\

Our results are very encouraging in the perspective of doing true non-perturbative calculations of bound state systems in a field theoretical framework.



\begin{thebibliography}{9}
\bibitem{bpp}
S.J.~Brodsky, H.C.~Pauli and S.S.~Pinsky, Phys. Rep. {\bf 301}
(1998) 299.
\bibitem{cdkm}
J.~Carbonell, B.~Desplanques, V.A.~Karmanov and J.-F.~Mathiot,
Phys. Rep. {\bf 300} (1998) 215.
\bibitem{kms}
V.A.~Karmanov, J.-F.~Mathiot and A.V.~Smirnov, Phys. Rev. {\bf
D69} (2004) 045009.
\bibitem{bckm}
D.~Bernard, Th.~Cousin, V.A.~Karmanov and J.-F.~Mathiot,  Phys.
Rev. {\bf D65} (2001) 025016.
\bibitem{Br_etal}
S.J.~Brodsky, V.A.~Franke, J.R.~Hiller, G.~McCartor, S.A.~Paston,
E.V.~Prokhvatilov, Nucl. Phys. {\bf 703} (2004) 333.
\bibitem{BHMc}
S.J.~Brodsky, J.R.~Hiller, G.~McCartor, arXiv:hep-ph/0508295.
\bibitem{mks}
J.-F.~Mathiot, V.A.~Karmanov and A.V.~Smirnov, Few-Body Systems
{\bf 36} (2005) 173.
\bibitem{km}
V.A.~Karmanov and J.-F.~Mathiot, Nucl. Phys. {\bf A602} (1996)
388.
\bibitem{dkm}
J.-J.~Dugne, V.A.~Karmanov and J.-F.~Mathiot, Eur. Phys. J. {\bf
C22} (2001) 105.
\end{thebibliography}
\end{document}